# Fast Matrix Rank Algorithms and Applications

Ho Yee Cheung*, Tsz Chiu Kwok, Lap Chi Lau

The Chinese University of Hong Kong

April 3, 2012

**Abstract**

We consider the problem of computing the rank of an $m \times n$ matrix $A$ over a field. We present a randomized algorithm to find a set of $r = \operatorname{rank}(A)$ linearly independent columns in $\tilde{O}(|A| + r^\omega)$ field operations, where $|A|$ denotes the number of nonzero entries in $A$ and $\omega < 2.38$ is the matrix multiplication exponent. Previously the best known algorithm to find a set of $r$ linearly independent columns is by Gaussian elimination, with running time $O(mnr^{\omega-2})$. Our algorithm is faster when $r < \max\{m,n\}$, for instance when the matrix is rectangular. We also consider the problem of computing the rank of a matrix dynamically, supporting the operations of rank one updates and additions and deletions of rows and columns. We present an algorithm that updates the rank in $\tilde{O}(mn)$ field operations. We show that these algorithms can be used to obtain faster algorithms for various problems in numerical linear algebra, combinatorial optimization and dynamic data structure.

## 1 Introduction

Given an $m \times n$ matrix $A$ over a field $F$, the rank of $A$, denoted by $\operatorname{rank}(A)$, is the maximum number of linearly independent columns of $A$. We consider the problem of computing $\operatorname{rank}(A)$ and finding a set of $\operatorname{rank}(A)$ linearly independent columns efficiently. It is a basic computational problem in numerical linear algebra that is used as a subroutine for other problems [40, 20]. It also has a number of applications in graph algorithms and combinatorial optimization: Some of the fastest algorithms for graph matching [33, 22], graph connectivity [10, 35, 12], matroid optimization problems [22, 11] are based on fast algorithms for computing matrix rank and finding linearly independent columns.

The traditional approach to compute $\operatorname{rank}(A)$ is by Gaussian elimination. For an $m \times n$ matrix with $m \leq n$, it is known that this approach can be implemented in $O(nm^{\omega-1})$ field operations [6, 27], where $\omega < 2.38$ is the matrix multiplication exponent [14]. More generally, given an $m \times n$ matrix and a parameter $k \leq m \leq n$, one can compute $\min\{\operatorname{rank}(A), k\}$ in $O(nmk^{\omega-2})$ field operations [39]. The Gaussian elimination approach has the advantage that it can also find a set of $\min\{\operatorname{rank}(A), k\}$ linearly independent columns in the same time. These algorithms are deterministic.

There are also randomized algorithms to compute the value of $\operatorname{rank}(A)$ more efficiently. We know of three approaches.

1. The first approach is to do an efficient preconditioning [29, 9]. Let $B = T_1 A T_2$ where $T_1$ and $T_2$ are Toeplitz matrices with entries chosen uniformly and randomly from a large enough subset of the field.

---

*Now at University of Southern California.

Then $B$ can be computed in $\tilde{O}(mn)$ time because of the structure of $T_1$ and $T_2$. Let $r = \text{rank}(A)$. It is proven that [29] the leading $r \times r$ minor of $B$ is of full rank with high probability. Thus $\text{rank}(A)$ can be computed in $\tilde{O}(mn + r^\omega)$ field operations. There is another efficient preconditioner based on butterfly network [9] with similar property and running time. This approach works for any field.

2. There is a black-box approach that computes $\text{rank}(A)$ in $O(m \cdot |A|)$ field operations [43, 20, 37] where $|A|$ is the number of nonzero entries of $A$. The method is based on computing the minimal polynomial of $A$ for Krylov subspaces. It does not require to store $A$ explicitly, as long as there is an oracle to compute $Ab$ for any vector $b$. This approach is fast when the matrix is sparse, and it works for any field.

3. Another approach is based on random projection for matrices over real numbers. Given an $m \times n$ matrix $A$ over $\mathbb{R}$, one can reduce $A$ into an $m \times (m \log m)$ matrix $A'$ so that $\text{rank}(A) = \text{rank}(A')$ with high probability [36] by the Johnson-Lindenstrauss lemma. The matrix $A'$ can be computed efficiently using fast Johnson-Lindenstrauss transform [2, 3], and this implies an $\tilde{O}(nm + m^\omega)$ randomized algorithm to compute $\text{rank}(A)$. This approach is only known to work for matrices over real numbers.

We remark that only the Gaussian elimination approach can also find a set of $\text{rank}(A)$ linearly independent columns, while other approaches can only compute the value of $\text{rank}(A)$.

## 1.1 Main Results

We present faster randomized algorithms to compute matrix rank and show their applications. In this section we use the $\tilde{O}$ notation to hide (small) polylog factors in the time bounds. We will state the precise time bounds in the technical sections. We assume that there is at least one nonzero entry in each row and each column, and thus $|A| \geq \max\{m, n\}$.

**Theorem 1.1.** *Given an $m \times n$ matrix $A$ over a field $F$ and a parameter $k$ where $k \leq m \leq n$, there is a randomized algorithm to compute $\min\{\text{rank}(A), k\}$ in $O(|A| + \min\{k^\omega, k|A|\})$ field operations where $|A|$ denotes the number of nonzeros in $A$. Furthermore, there is a randomized algorithm to find a set of $\min\{\text{rank}(A), k\}$ linearly independent columns in $\tilde{O}(|A| + k^\omega)$ field operations.*

For computing $\min\{\text{rank}(A), k\}$, previous algorithms require $\tilde{O}(mn + k^\omega)$ field operations, while we replace the $mn$ term by $|A|$ and remove the (small) polylog factor. More importantly, we can also find a set of $\min\{\text{rank}(A), k\}$ linearly independent columns in about the same time, which is considerably faster than the $O(mnk^{\omega-2})$ algorithm by Gaussian elimination when $k$ is small. For instances, we can find a set of $k = n^{1/\omega} \approx n^{0.42}$ linearly independent columns in $\tilde{O}(|A|)$ field operations, and a set of $k = n^{1/(\omega-1)} \approx n^{0.72}$ linearly independent columns in $\tilde{O}(mn)$ field operations, while previously it was possible only for $k = O(\text{polylog}(n))$. The algorithm for finding linearly independent columns is needed in all applications of Theorem 1.1 that we will describe in the next subsection.

We also present a dynamic algorithm to efficiently update the matrix rank.

**Theorem 1.2.** *Given an $m \times n$ matrix $A$ over a field $F$, there is a randomized algorithm to compute $\text{rank}(A)$ dynamically in $\tilde{O}(mn)$ field operations in the worst case, supporting the operations of rank one updates and adding and deleting rows and columns.*

Previously there is a dynamic algorithm to update the matrix rank in $O(n^2)$ field operations for an $n \times n$ square matrix, supporting the operation of rank one updates [16, 35]. There are also subquadratic dynamic algorithms to update the matrix rank when few entries are changed [16, 35]. Our algorithm supports the new operations of adding and deleting rows and columns. These new operations will be useful in computing graph connectivities dynamically (see Theorem 1.5).

|  | graph matching | linear matroid intersection | linear matroid union |
|---|---|---|---|
| combinatorial | $O(\sqrt{\mathsf{opt}} \cdot |E|)$ [32, 21] | $\tilde{O}(nr(\mathsf{opt})^{\frac{1}{4-\omega}})$ [19] | $\tilde{O}(nrb(\mathsf{opt}) + nb^2(\mathsf{opt})^2)$ [15] |
| algebraic | $O(|V|^\omega)$ [33] | $O(nr^{\omega-1})$ [22] | − |
| this paper | $\tilde{O}(|E| + (\mathsf{opt})^\omega)$ | $\tilde{O}(nr + n(\mathsf{opt})^{\omega-1})$ | $\tilde{O}(nr(\mathsf{opt}) + b^3(\mathsf{opt})^3)$ |

Table 1.1: Time complexity of algorithms for some problems in combinatorial optimization

## 1.2 Applications

The matrix rank algorithms can be readily applied to various problems in numerical linear algebra, combinatorial optimization, and dynamic data structure. First we show that the algorithms can be applied to computing a rank-one decomposition, finding a basis of the null space, and performing matrix multiplication for a low rank matrix.

**Theorem 1.3.** *Let $A$ be an $m \times n$ matrix over a field $F$. Let $r = \text{rank}(A)$. Let $m' = \min\{m, n\}$. Let $\omega(a, b, c)$ be the exponent for multiplying an $n^a \times n^b$ matrix with an $n^b \times n^c$ matrix.*

1. *There is a randomized algorithm to compute an $m \times r$ matrix $X$ and an $r \times n$ matrix $Y$ such that $A = XY$ in $\tilde{O}(|A| + r^{\omega(1,1,\log_r m')}) = \tilde{O}(|A| + m'r^{\omega-1})$ steps.*

2. *There is a randomized algorithm to find a basis of the null space of $A$ in $\tilde{O}(|A| + r^{\omega(1,1,\log_r n)}) = \tilde{O}(|A| + nr^{\omega-1})$ steps.*

3. *Let $A$ and $B$ be $n \times n$ matrices. There is a randomized algorithm to compute $AB$ in $\tilde{O}(n^{\omega(\log_n r, 1, 1)}) = \tilde{O}(n^2 r^{\omega-2})$ steps.*

*The success probability for all three tasks is at least $1 - O(\log(nm)/|A|^{1/3})$.*

Previously the best known algorithms require $\tilde{\Theta}(mnr^{\omega-2})$ for the first two tasks, and $\tilde{\Theta}(n^2 r^{\omega-2})$ for the third task. Our algorithms are faster than the existing algorithms, especially when $r$ is small. For rank-one decomposition, the algorithm requires only $\tilde{O}(mn)$ field operations when $r \leq (\max\{m, n\})^{0.72}$. For finding null space, the algorithm requires only $\tilde{O}(mn)$ field operations when $r \leq m^{0.72}$. For matrix multiplication, the algorithm requires only $\tilde{O}(n^{2+\epsilon})$ field operations when $r \leq n^{0.29}$, since $\omega(0.29, 1, 1) \leq 2 + \epsilon$ for any $\epsilon > 0$ [13, 26]. The statement about matrix multiplication essentially says that the problem of multiplying two $n \times n$ matrices while one matrix is of rank $r$ can be reduced to the problem of multiplying an $r \times n$ matrix and an $n \times n$ matrix.

In combinatorial optimization, there are algebraic formulations of the problems that relate the optimal value to the rank of an associated matrix. Using this connection, we can apply the algorithm in Theorem 1.1 to obtain fast algorithms for graph matching and matroid optimization problems. See Section 4 for the definitions of these problems.

**Theorem 1.4.** *Let $\mathsf{opt}$ be the optimal value of an optimization problem.*

1. *Given an undirected graph $G = (V, E)$, there is a randomized algorithm to find a matching of size $\min\{\mathsf{opt}, k\}$ in $\tilde{O}(|E| + k^\omega)$ time.*

2. *Given a linear matroid intersection problem or a linear matroid parity problem with an $r \times 2n$ matrix $A$, there is a randomized algorithm to find a solution of size $\min\{\mathsf{opt}, k\}$ in $\tilde{O}(|A| + nk^{\omega-1})$ time.*

3. *Given a linear matroid union problem with an $r \times n$ matrix $|A|$, there is a randomized algorithm to find $\min\{\mathsf{opt}, k\}$ disjoint bases in $\tilde{O}(k|A| + \min\{k^{\omega+1}b^\omega, k^3 b^3\})$ time, where $b$ denotes the size of a basis.*

Table 1.1 lists the time complexity of the best known combinatorial algorithms and algebraic algorithms for these problems. Notice that previous algebraic algorithms have the same time complexity even when the optimal value is small. On the other hand, combinatorial algorithms for these problems are based on finding augmenting structures iteratively, and thus the number of iterations and the overall complexity depend on the optimal value. While the previous algebraic algorithms are faster than combinatorial algorithms only when the optimal value is large, the results in Theorem 1.4 show that the algebraic approach can be faster for any optimal value. For the matroid optimization problems, the algorithms in Theorem 1.4 are faster than previous algorithms in any setting. The result in the graph matching problem can be applied to the subset matching problem [4] and the lopsided bipartite matching problem [8]. See Section 4 for more discussions on previous work for these problems.

The dynamic matrix rank algorithm in Theorem 1.2 can be applied to obtain a dynamic algorithm to compute edge connectivities in a directed graph.

**Theorem 1.5.** *Given an uncapacitated directed graph $G = (V, E)$, there is a randomized algorithm to compute all-pairs edge-connectivities dynamically in $\tilde{O}(|E|^2)$ time and $\tilde{O}(|E|^2)$ space, supporting the operations of adding and deleting edges.*

In undirected graphs, there are polylogarithmic time dynamic algorithms for computing $k$-edge-connectivity for $k \leq 2$ [24], and a $\tilde{O}(|V||E|)$ time algorithm to compute all pairs edge connectivities [5]. The corresponding problems are more difficult for directed graphs. There is a subquadratic dynamic algorithm for computing 1-edge-connectivity in directed graphs [34]. For all pairs edge connectivities in directed graphs, we do not know of any dynamic algorithm that is faster than the straightforward dynamic algorithm that uses $\Theta(|V|^2|E|)$ time and $\Theta(|V|^2|E|)$ space, by storing the flow paths for each pair and running an augmentation step for each pair after each edge update. For graphs with $O(|V|)$ edges (e.g. planar graphs), the amortized complexity of our algorithm to update the edge connectivity for one pair is $O(1)$ field operations.

## 1.3 Methods

Similar to the preconditioning approach, our approach is to compress the matrix $A$ into a $O(k) \times O(k)$ matrix $B$, while keeping the property that $\min\{\text{rank}(B), k\} = \min\{\text{rank}(A), k\}$ with high probability. To illustrate the ideas, we consider the special case of computing the rank of a rectangular $m \times n$ matrix $A$ where $m < n$, and the goal is to compress the matrix into an $m \times O(m)$ matrix with $\text{rank}(B) = \text{rank}(A)$. We present two efficient methods to do the compression, assuming the field size is sufficiently large for now. The first method is inspired by the random linear coding algorithm [23] in network coding [1] and its efficient implementation using superconcentrators [12]. Suppose we write each column of $B$ as a random linear combination of all the columns of $A$. Then it can be shown that $\text{rank}(B) = \text{rank}(A)$ with high probability by the Schwartz-Zippel lemma, but the direct implementation of this method requires a fast rectangular matrix multiplication algorithm. To do the compression efficiently, we use a construction similar to that of magical graphs [25] in the construction of superconcentrators. We prove that if each column of $B$ is a random linear combination of $O(n/m)$ random columns of $A$, it still holds with high probability that $\text{rank}(B) = \text{rank}(A)$. In addition, this property still holds when each column of $A$ is involved in only $O(1)$ linear combinations, and so the sparsity of the matrix can be preserved, i.e. $|B| = O(|A|)$. Hence $B$ can be constructed in $O(|A|)$ field operations, and $\text{rank}(B)$ can be computed in $O(m^\omega)$ field operations, and thus $\text{rank}(A)$ can be computed in $O(|A| + m^\omega)$ field operations. Based on a bounded degree condition of the magical graphs, the above procedure can be applied iteratively to reduce the number of columns of $A$ progressively, so that a set of $\text{rank}(A)$ linearly independent columns in $A$ can be found in $\tilde{O}(|A| + m^\omega)$ field operations.

Another method to compute $B$ is to multiply $A$ with an $n \times m$ random Vandermonde matrix $V$ with only one variable. We show that $\text{rank}(B) = \text{rank}(A)$ with high probability, by using the Cauchy-Binet formula and a base exchange argument. The $m \times m$ matrix $B = AV$ can be computed in $\tilde{O}(mn)$ field operations using fast Fourier transform. This provides an alternative way to compute rank$(A)$ in $\tilde{O}(mn + m^\omega)$ field operations, although it is slower than the above method. The advantage of this method is that it allows us to update the matrix rank efficiently when we add and delete rows and columns of $A$, because of the special structures of the Vandermonde matrices. For instance, when the $m \times n$ matrix $A$ is changed from $m < n$ to $m > n$, we can change the representation from $B = AV$ to $B' = V'A$ by doing an inverse Fourier transform. This allows us to update rank$(A)$ in $\tilde{O}(mn)$ field operations in the worst case.

## 2  Fast Matrix Rank Algorithms

In this section we will prove Theorem 1.1. Let $A$ be an $m \times n$ matrix over a field $F$. We will assume that $A$ is given by a list of the value and the position of its non-zero entries, and each row and column of $A$ contains at least one non-zero entry, so $|A| \geq \max(n, m)$. We will also assume that $|F| = \Omega(n^4)$ by the following lemma using standard techniques.

**Lemma 2.1.** *Let $A$ be an $m \times n$ matrix over a field $F$ with $p^c$ elements. We can construct a finite field $F'$ with $p^{ck} = \Omega(n^4)$ elements and an injective mapping $f : F \to F'$ so that the image of $F$ is a subfield of $F'$. Then the $m \times n$ matrix $A'$ where $a'_{ij} = f(a_{ij})$ satisfies the property that* rank$(A')$ = rank$(A)$. *This preprocessing step can be done in $O(|A|)$ field operations. Each field operation in $F'$ can be done in $\tilde{O}(\log|F| + \log n)$ steps.*

*Proof.* All the statements in this proof refer to the statements in book [20]. Let $q = p^c$. By Theorem 14.42 in [20], we can construct a monic irreducible polynomial $h$ with degree $k$ in expected $O(k^2 \log^2 k \log\log k(\log k + \log q))$ field operations in $F_q$. Note that the collection of polynomials with coefficients in $F_q$ and degree less than $k$, with multiplications and division under modulo $h$, is a field with size $q^k$. So we can use an ordered $k$-tuple $(c_0, c_1, \ldots, c_{k-1})$ with $c_i \in F_q$ to represent an element $\sum_{i=0}^{k-1} c_i x^i$ in $F_{q^k}$. The injective mapping $f$ in the statement is just the identity mapping in this construction, i.e. $f(c) = (c, 0, 0, \ldots, 0)$. The overall preprocessing time is $O(|A| + k^2 \log^2 k \log\log k(\log k + \log q)) = O(|A|)$ field operations in $F_{q^k}$. It follows directly that rank$(A')$ = rank$(A)$.

Additions and subtractions are done coordinate-wise, and thus requires $O(k)$ field operations. For two polynomials $g_1$ and $g_2$ with coefficients in $F_q$ and degree less than $k$, $g_1 \times g_2$ can be computed in $O(k \log k \log\log k)$ field operations in $F_q$, by Theorem 8.22 and Exercise 8.30 in [20]. So $g_1 \times g_2 \mod h$ can be computed in $O(k \log k \log\log k)$ field operations in $F_q$ by Theorem 9.6 in [20]. Division $a/b$ is done by multiplying the inverse $a \times b^{-1}$. The inverse $b^{-1}$ can be computed by the extended euclidean algorithm, in $O(k \log^2 k \log\log k)$ field operations in $F_q$ by Theorem 11.7 in [20]. Since field operations in $F_q$ can be computed in $\tilde{O}(\log q)$ steps, the operations in $F_{q^k}$ in our representation can be done in $\tilde{O}(\log q^k)$, where $\tilde{O}$ hides some polylog factors of $\log q^k$. □

Suppose a parameter $k$ is given and the task is to compute $\min\{\text{rank}(A), k\}$. Our approach is to compress the matrix into a $O(k) \times O(k)$ matrix whose rank is at least $\min\{\text{rank}(A), k\}$ with high probability. Our method is inspired by the random linear coding algorithm [23, 12] in network coding [1]. We can construct an $m \times k$ matrix $B$ where each column of $B$ is a random linear combination of the columns of $A$, i.e. $B_i = \sum_{j=1}^{n} c_{j,i} A_j$ where $A_i$ and $B_i$ denote the $i$-th column of $A$ and $B$ respectively and $c_{j,i}$ is uniformly independent random element in $F$. In other words, $B = AC$ where $C$ is an $n \times k$ matrix where each entry is a random element in $F$. It can be shown that rank$(B)$ = rank$(A)$ with high probability using the Schwartz-Zippel lemma (see Lemma 2.3), but it requires a fast rectangular matrix multiplication algorithm [26] to compute $B$. We observe that this way of constructing $B$ is the same as doing the random linear coding algorithm in a single vertex with $n$ incoming edges and $k$ outgoing edges. And so the ideas of using a superconcentrator to do the random linear coding efficiently [12] can be applied to construct an $m \times k$ matrix $B'$ in $O(mn)$ field operations, while $\text{rank}(B') = \text{rank}(A)$ with high probability (see Appendix A for details). This implies that $\min\{\text{rank}(A), k\}$ can be computed in $O(mn + k^\omega)$ field operations with high probability, improving the existing algorithms by removing the polylog factor. There are, however, two disadvantages of this method. One is that the compression algorithm requires $\Theta(mn)$ field operations even when $A$ is a sparse matrix. Another is that we do not know how to find a set of $\min\{\text{rank}(A), k\}$ linearly independent columns of $A$ using this method.

To improve the compression algorithm, we use sparse random bipartite graphs similar to that of magical graphs [25] in the construction of superconcentrators. The idea is to choose a sparse matrix $C$ so that $B = AC$ can be computed efficiently while $\min\{\text{rank}(B), k\} = \min\{\text{rank}(A), k\}$ with high probability, but it is easier to explain the compression algorithm using graph theoretical concepts. Our construction requires a probability distribution of bipartite graphs with the following properties.

**Definition 2.2** (Magical Graph). *A random bipartite graph $G = (X, Y; E)$ is $(k, \epsilon)$-magical if for every subset $S \subseteq X$ with $|S| = k$, the probability that there is a matching in $G$ in which every vertex in $S$ is matched is at least $1 - \epsilon$.*

Notice that this definition only requires a particular subset $S$ of size $k$ can be matched to the other side with high probability, while the definition in [25] requires that all subsets up to certain size can be matched to the other side. In the following we first see how to use a magical graph to do compression (Figure 2.1), and then we will see how to generate a magical graph with good parameters efficiently.

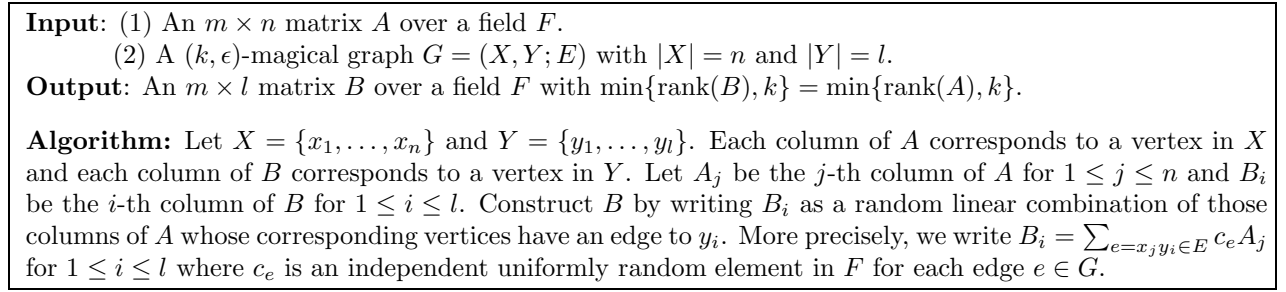

**Input**: (1) An $m \times n$ matrix $A$ over a field $F$.
  (2) A $(k, \epsilon)$-magical graph $G = (X, Y; E)$ with $|X| = n$ and $|Y| = l$.
**Output**: An $m \times l$ matrix $B$ over a field $F$ with $\min\{\text{rank}(B), k\} = \min\{\text{rank}(A), k\}$.

**Algorithm:** Let $X = \{x_1, \ldots, x_n\}$ and $Y = \{y_1, \ldots, y_l\}$. Each column of $A$ corresponds to a vertex in $X$ and each column of $B$ corresponds to a vertex in $Y$. Let $A_j$ be the $j$-th column of $A$ for $1 \leq j \leq n$ and $B_i$ be the $i$-th column of $B$ for $1 \leq i \leq l$. Construct $B$ by writing $B_i$ as a random linear combination of those columns of $A$ whose corresponding vertices have an edge to $y_i$. More precisely, we write $B_i = \sum_{e = x_j y_i \in E} c_e A_j$ for $1 \leq i \leq l$ where $c_e$ is an independent uniformly random element in $F$ for each edge $e \in G$.

Figure 2.1: Compression Algorithm by Magical Graphs.

**Lemma 2.3.** *The probability that the algorithm in Figure 2.1 returns a matrix $B$ such that $\min\{\text{rank}(B), k\} = \min\{\text{rank}(A), k\}$ is at least $1 - \epsilon - k/|F|$.*

*Proof.* Clearly $\text{rank}(B) \leq \text{rank}(A)$ since the column space of $B$ is a subspace of the column space of $A$. So $\min\{\text{rank}(B), k\} \leq \min\{\text{rank}(A), k\}$, and it remains to show that $\text{rank}(B) \geq \min\{\text{rank}(A), k\}$ with high probability.

Let $k' = \min\{\text{rank}(A), k\}$. Let $S$ be a set of linearly independent columns of $A$ with $|S| = k'$, and let $A_{U,S}$ be a $k' \times k'$ submatrix of $A$ with $\text{rank}(A_{U,S}) = k'$. Since $G$ is a magical graph, the probability that there is a matching $M$ in which every vertex in $S$ is matched is at least $1 - \epsilon$. Suppose such a matching $M$ exists and let $T$ be the neighbors of $S$ in $M$ with $|T| = |S| = k'$. If we view each $c_e$ as a variable, then $\det(B_{U,T})$ is a multivariate polynomial with total degree $k'$. By setting $c_e = 1$ for each $e \in M$ and $c_e = 0$ for each $e \in E - M$, we get that $B_{U,T} = A_{U,S}$ and thus $\det(B_{U,T})$ is a nonzero multivariate polynomial since $A_{U,S}$ is of full rank. By the Schwartz-Zippel lemma, if we substitute each variable $c_e$ by a random element in a field $F$, then the probability that $\det(B_{U,T}) = 0$ is at most $k'/|F| \leq k/|F|$. So, if $G$ has a matching that matches every vertex in $S$, then $\text{rank}(B) \geq \text{rank}(B_{U,T}) = k'$ with probability at least $1 - k/|F|$. Therefore the algorithm succeeds with probability at least $1 - \epsilon - k/|F|$. □

We show that a magical graph with good parameters can be generated efficiently.

**Lemma 2.4.** *There is a randomized $O(|X|)$ time algorithm to construct a $(k, O(1/k))$-magical graph $G = (X, Y; E)$ for any values of $|X| \geq |Y| \geq ck$ where $c \geq 11$, with the additional properties that each vertex of $X$ is of degree 2 and each vertex of $Y$ is of degree at most $2\lceil |X|/|Y| \rceil$.*

We note that the magical graphs in [25] cannot be used directly because of the following reasons: (1) the failure probability in [25] is a constant while we need a much smaller failure probability in order to find a set of linearly independent columns, (2) we need the additional property that the graph is almost regular to find a set of linearly independent columns. Also the proof is somewhat different and the degree of the vertices in $X$ is smaller.

*Proof.* The generation algorithm is simple. We assume that $|X|$ is a multiple of $|Y|$; otherwise we construct a slightly larger graph and delete the extra vertices. We first construct a 2-regular graph $G'$ with $|X|$ vertices on both sides, by taking the union of two random perfect matchings independently from $|X|$ vertices to $|X|$ vertices. Then we divide the $|X|$ vertices on one side into $|Y|$ groups where each group has $|X|/|Y|$ vertices. We obtain $G$ by merging each group into a single vertex, and so each vertex in $Y$ is of degree $2|X|/|Y|$.

For any $S \subseteq X$ with $|S| = k$, we analyze the probability that there is a matching in $G$ in which every vertex in $S$ is matched. By Hall's theorem, we need to show that for any $S' \subseteq S$, the neighbor set of $S'$ in $G$ is of size at least $|S'|$. To analyze the probability that the neighbor set of $S'$ is at least $|S'|$ for a fixed $S' \subseteq S$, we consider the equivalent random process where the $2|S'|$ edges incident on $S'$ are added one by one. Consider the $i$-th edge added. We say that it is a bad edge if the other endpoint falls in the same group with some previously added edges. If the neighbor set size of $S'$ is less than $|S'|$, then there must be at least $|S'| + 1$ bad edges out of the $2|S'|$ edges, and the probability that an edge is bad is less than $|S'|/|Y|$. So the probability that the neighbor set size of $S'$ is less than $|S'|$ is less than

$$\binom{2|S'|}{|S'|+1} \times \left(\frac{|S'|}{|Y|}\right)^{|S'|+1}$$

by a union bound on the possible $|S'| + 1$ bad edges. Summing over the choices of the size of $S'$ and the choices of $S'$ with that size, we have that the probability that there is a subset $S' \subseteq S$ with less than $|S'|$ neighbors is at most

$$\sum_{z=0}^{k} \binom{2z}{z+1} \left(\frac{z}{|Y|}\right)^{z+1} \binom{k}{z} \leq \sum_{z=0}^{k} 2^{2z} \left(\frac{z}{|Y|}\right)^{z+1} \left(\frac{ke}{z}\right)^{z}$$

$$\leq \sum_{z=0}^{k} \left(\frac{4e}{c}\right)^{z} \frac{z}{ck}$$

$$= O(1/k),$$

using $|Y| \geq ck$ and the identity $\sum_{z=0}^{\infty} r^{z} \cdot z = r/(1-r)^{2}$ for $r < 1$, and setting $r = 4e/c$ as $c \geq 11 > 4e$ by our assumption. Therefore, by Hall's theorem, the probability that there is a matching in which every vertex in $S$ is matched is at least $1 - O(1/k)$. □

We can combine Lemma 2.4 and Lemma 2.3 to obtain an efficient compression algorithm.

**Theorem 2.5.** *Suppose an $m \times n$ matrix $A$ over a field $F$ is given. Given $k$, there is an algorithm that constructs an $m \times O(k)$ matrix $B$ over $F$ with the following properties.*

1. $\min\{\text{rank}(A), k\} = \min\{\text{rank}(B), k\}$ *with probability at least $1 - O(1/k) - O(k/|F|)$.*

2. $|B| = O(|A|)$ and $B$ can be constructed in $O(|A|)$ field operations.

*Proof.* We can assume $n \geq 11k$; otherwise we can just let $B = A$. We construct a $(k, O(1/k))$-magical graph $G = (X, Y; E)$ with $|X| = n$ and $|Y| = 11k$ in $O(n)$ time by Lemma 2.4, with the additional property that each vertex in $X$ is of degree two. We use $G$ in the algorithm in Figure 2.1 to obtain an $m \times 11k$ matrix $B$ over $F$. Since each vertex of $X$ is of degree two, each entry of $A$ is related to two entries in $B$. We can represent $B$ by listing the value and position of its nonzero entries without handling duplicate positions, i.e. each nonzero entry in $A$ introduces exactly two entries in $B$. Therefore, $|B| = 2|A|$ and $B$ can be constructed in $O(|A|)$ field operations. The probability that $\min\{\text{rank}(A), k\} = \min\{\text{rank}(B), k\}$ is at least $1 - O(1/k) - O(k/|F|)$ by Lemma 2.3. □

The first part of Theorem 1.1 follows.

**Theorem 2.6.** *Suppose an $m \times n$ matrix $A$ over a field $F$ is given with $m \leq n$. There is an algorithm to compute $\min\{\text{rank}(A), k\}$ for a given $k \leq m$ in $O(|A| + \min\{k^\omega, k|A|\})$ field operations with success probability at least $1 - O(1/n^{1/3})$. Furthermore, there is an algorithm to compute $r = \text{rank}(A)$ in $O(|A| \log r + \min\{r^\omega, r|A|\})$ field operations with success probability $1 - O(1/n^{1/3})$. Each field operation can be done in $\tilde{O}(\log n + \log |F|)$ steps.*

*Proof.* We can assume that $|F| = \Omega(n^4)$ by Lemma 2.1. We also assume that $k \geq n^{1/3}$; otherwise if $k < n^{1/3}$ we just reset $k$ to be $n^{1/3}$. We apply Theorem 2.5 to compress the $m \times n$ matrix $A$ into an $m \times O(k)$ matrix $B$. Then $\min\{\text{rank}(B), k\} = \min\{\text{rank}(A), k\}$ with probability at least $1 - O(1/k) - O(k/|F|) = 1 - O(1/n^{1/3})$ since $n^{1/3} \leq k \leq n$ and $|F| = \Omega(n^4)$. And $B$ can be constructed in $O(|A|)$ field operations and $|B| = O(|A|)$. We then apply Theorem 2.5 again on $B^T$ to compress the $m \times O(k)$ matrix $B$ into an $O(k) \times O(k)$ matrix $C$. Then $\min\{\text{rank}(C), k\} = \min\{\text{rank}(B), k\}$ with probability at least $1 - O(1/n^{1/3})$ and $C$ can be constructed in $O(|A|)$ field operations with $|C| = O(|A|)$. Now we can compute $\text{rank}(C)$ in $O(k^\omega)$ field operations by using fast matrix multiplication [6]. Alternatively, we can compute $\text{rank}(C)$ in $O(k|C|) = O(k|A|)$ field operations using the black box approach in [37, 20]. Thus $\min\{\text{rank}(A), k\}$ can be computed in $O(|A| + \min\{k^\omega, k|A|\})$ field operations with success probability $1 - O(1/n^{1/3})$, since $k^\omega \leq n \leq |A|$ for $k \leq n^{1/3}$. To compute $\text{rank}(A)$, we can simply apply the above algorithm with $k = n^{1/3}, 2n^{1/3}, 4n^{1/3}, \ldots, 2^{\log n^{2/3}} n^{1/3}$ until the algorithm returns an answer smaller than $k$ or $A$ is of full rank. Let $r = \text{rank}(A)$. The failure probability is bounded by $O(1/n^{1/3})$ since sum of $1/k$ is less than $2/n^{1/3}$. The number of field operations needed is $O(|A| \log r + \min\{r^\omega, r|A|\})$ since sum of $k^\omega$ is $O(r^\omega)$ and sum of $k|A|$ is $O(r|A|)$. If the field size is $|F|$, then each field operation can be done in $\tilde{O}(\log |F|)$ steps using fast arithmetic algorithms by Lemma 2.1. Since we assume $|F| = \Omega(n^4)$, each field operation can be done in $\tilde{O}(\log \max\{|F|, n^4\}) = \tilde{O}(\log |F| + \log n)$ steps. □

We can improve Theorem 2.6 slightly to reduce the time complexity to $O(\min\{|A| \log r, nm\} + \min\{r^\omega, r|A|\})$ field operations. This is done by computing the compressed matrices aggregately and we omit the details here (see Section A for such a statement using superconcentrators).

Finally, we find a set of $\min\{\text{rank}(A), k\}$ linearly independent columns of $A$ by applying the compression algorithm iteratively to reduce the number of columns of $A$ progressively. The bounded degree condition of magical graphs is important in the following algorithm.

**Theorem 2.7.** *Suppose an $m \times n$ matrix $A$ over a field $F$ is given. There is an algorithm to find a set of $\min\{\text{rank}(A), k\}$ linearly independent columns of $A$ for a given $k$ in $O((|A| + k^\omega) \log n)$ field operations with success probability at least $1 - O((\log n)/n^{1/3})$, while each field operation can be done in $\tilde{O}(\log n + \log |F|)$ steps.*

*Proof.* The idea is to compress the matrix $A$ to a smaller matrix $B$, and then we can focus on columns in $A$ that corresponds to a set of linearly independent columns in $B$. We will show that this reduce the number of columns of $A$ by a constant factor, so that we can repeat this procedure to reduce the number of columns of $A$ to a small number.

We assume $k \geq n^{1/3}$ as in Theorem 2.6. Let $c = 11$. If $m > ck$, we first apply the algorithm in Theorem 2.5 to $A^T$ to compress $A$ into a $ck \times n$ matrix $A'$ in $O(|A|)$ field operations. We claim that if a set of columns is linearly independent in $A$ then it is linearly independent in $A'$ with high probability. Let $S$ be a set of linearly independent columns of $A$. By Theorem 2.5 we have $\text{rank}(A'_{[ck],S}) = \text{rank}(A_{[m],S}) = |S|$ with probability at least $1 - O(1/n^{1/3})$, and thus $S$ is a set of linearly independent columns in $A'$. Henceforth we use the smaller matrix $A'$ to find the linearly independent columns of $A$.

We then use the algorithm in Theorem 2.5 to compress $A'$ into a $ck \times ck$ matrix $B$ in $O(|A'|) = O(|A|)$ field operations, while $\min\{\text{rank}(A), k\} = \min\{\text{rank}(B), k\}$ with probability at least $1 - O(1/n^{1/3})$. Since $B$ is a $ck \times ck$ matrix, we can directly find a set $S$ of $\min\{\text{rank}(B), k\}$ linearly independent columns in $B$ in $O(k^\omega)$ field operations using fast matrix multiplication [6]. Let $G = (X, Y; E)$ be the magical graph used in the compression algorithm with $|X| = n$ and $|Y| = ck$. Let $T$ be the set of columns in $A$ that correspond to the neighbors of the vertices corresponding to $S$ in $G$. By the bounded degree condition of $G$, each vertex corresponding to a column in $S$ is of degree at most $2|X|/|Y| = 2n/(ck)$ and hence $|T| \leq 2n|S|/(ck) \leq 2n/c < n/5$. Observe that the $ck \times |T|$ submatrix $A'_{R',T}$ of $A'$ is of rank at least $\min\{\text{rank}(A), k\}$, since the column space of $S$ in $B$ is spanned by the column space of $A'_{R',T}$. Thus we have reduced the original problem to finding a set of $\min\{\text{rank}(A'_{R',T}), k\}$ linearly independent columns in a $ck \times (n/5)$ matrix $A'_{R',T}$. We can repeat the above algorithm until the number of columns is reduced to $O(k)$. Since each time we can reduce the number of columns by a constant factor, we need to repeat the algorithm at most $O(\log n)$ times. So the whole algorithm can be done in at most $O((|A| + k^\omega) \log n)$ field operations, and the failure probability is at most $O((\log n)/n^{1/3})$. □

Theorem 1.1 follows from Theorem 2.6 and Theorem 2.7.

## 3 Dynamic Matrix Rank Algorithm

In this section we present a dynamic algorithm for computing matrix rank and prove Theorem 1.2. Given an $m \times n$ matrix $A$, we will first show that $\text{rank}(A) = \text{rank}(AV)$ with high probability for an $n \times m$ random Vandermonde matrix $V$ with one variable. Then we show that the special structure of $V$ can be used to update the matrix rank of $A$ efficiently.

**Lemma 3.1.** *Let $m \leq n$. Let $V$ be a $n \times m$ random Vandermonde matrix with one variable, i.e., $V_{ij} = x^{ij}$ for $1 \leq i \leq n, 1 \leq j \leq m$. Suppose $x$ is chosen uniformly randomly in $F$, then for any $m \times n$ matrix $A$ over $F$, we have $\text{rank}(A) = \text{rank}(AV)$ with probability at least $1 - O(nm^2/|F|)$.*

*Proof.* We will first prove the lemma when $A$ is of full rank. Suppose $A$ is of full rank, then there exist $m$ linearly independent columns. Let $\mathcal{B} = \{I \subseteq [n] \mid |I| = m, \det(A_{[m],I}) \neq 0\}$ be the set of subsets of indices whose columns are linearly independent. Then $\mathcal{B} \neq \emptyset$. By the Cauchy-Binet formula,

$$\det(AV) = \sum_{I \subseteq [n], |I|=m} \det(A_{[m],I}) \det(V_{I,[m]})$$
$$= \sum_{I \in \mathcal{B}} \det(A_{[m],I}) \det(V_{I,[m]}).$$

Now view $\det(V_{I,[m]})$ as a polynomial in $x$. Suppose $I = \{i_1, i_2, \ldots i_m\}$ with $i_1 < i_2 < \cdots < i_m$. Let $S_m$ be the set of permutations of $[m]$. Note that

$$\det(V_{I,[m]}) = \sum_{\pi \in S_m} \mathrm{sgn}(\pi) \prod_{k=1}^{m} V_{i_k, \pi_k}$$
$$= \sum_{\pi \in S_m} \mathrm{sgn}(\pi) \prod_{k=1}^{m} x^{i_k \cdot \pi_k}$$
$$= \sum_{\pi \in S_m} \mathrm{sgn}(\pi) x^{\sum_{k=1}^{m} i_k \cdot \pi_k}.$$

By the rearrangement inequality $\sum_{k=1}^{m} i_k \pi_k \leq \sum_{k=1}^{m} i_k \cdot k$, and the equality holds only when $\pi_k = k$ for all $k$. Therefore,

$$\deg(\det(V_{I,[m]})) = \sum_{k=1}^{m} i_k \cdot k. \tag{3.1}$$

Clearly $\deg(\det(AV)) \leq \max_{I \in \mathcal{B}} \deg(\det(V_{I,[m]}))$. We are going to show that the equality actually holds, by arguing that $\max_{I \in \mathcal{B}} \deg(\det(V_{I,[m]}))$ is attained by only one $I$. Suppose not, let $J \neq K$ be two sets in $\mathcal{B}$ satisfying $\deg(\det(V_{J,[m]})) = \deg(\det(V_{K,[m]})) = \max_{I \in \mathcal{B}} \deg(\det(V_{I,[m]}))$. Let $j = \min\{i \mid i \in (J-K) \cup (K-J)\}$, and without loss of generality assume $j \in J$. It is well know that the sets in $\mathcal{B}$ are the bases of a (linear) matroid [38]. Therefore, by the base exchange property of a matroid ([38], Theorem 39.6), there exists some $k \in K$ such that $(J-\{j\}) \cup \{k\} \in \mathcal{B}$. By the choice of $j$, we have $j < k$, and thus $\deg(\det(V_{(J-\{j\}) \cup \{k\},[m]})) > \deg(\det(V_{J,[m]}))$ by (3.1), contradicting the maximality of $J$. In particular, since $\mathcal{B} \neq \emptyset$, this implies that $\deg(\det(AV)) > 0$ and thus is a non-zero polynomial. And $\deg(\det(V_{I,[m]})) = \sum_{k=1}^{m} i_k \cdot k \leq nm^2$ for any $I$. Therefore $\det(AV)$ is a non-zero polynomial with total degree at most $nm^2$. By the Schwartz-Zippel lemma, by substituting $x$ with a random element in $F$, we have $\det(AV) \neq 0$ and thus $\mathrm{rank}(AV) = \mathrm{rank}(A)$ with probability at least $1 - O(nm^2/|F|)$.

In general let $\mathrm{rank}(A) = k$ and assume without loss of generality that the first $k$ rows of $A$ are linearly independent. Clearly, $\mathrm{rank}(AV) \leq \mathrm{rank}(A)$ as the column space of $AV$ is spanned by the column space of $A$. We prove that $\mathrm{rank}(AV) \geq \mathrm{rank}(A)$ with high probability. Let $A'$ be the $k \times n$ submatrix of $A$ consisting of the first $k$ rows of $A$, and $V'$ be the $n \times k$ submatrix of $V$ consisting of the first $k$ columns of $V$. Then by the above argument we have that $\det(A'V') \neq 0$ with probability at least $1 - O(nm^2/|F|)$. Observe that $A'V'$ is equal to the $k \times k$ submatrix $(AV)_{[k],[k]}$ of $AV$. Therefore, we have $\mathrm{rank}(AV) \geq \mathrm{rank}((AV)_{[k],[k]}) = k = \mathrm{rank}(A)$ with probability at least $1 - O(nm^2/|F|)$. □

The matrix $AV$ can be computed efficiently using fast arithmetic algorithms: The multiplication of one row of $A$ with $V$ is equivalent to the evaluation of a polynomial over $m$ points $(x, x^2, \ldots, x^m)$ and this can be implemented efficiently using the following result.

**Theorem 3.2** ([20] Corollary 10.8). *There exist an algorithm that evaluates a degree $n$ polynomial $f \in F[x]$ at $m$ points in $F$, and it takes $O(n \log n \log \log n \log m)$ field operations.*

Therefore, the matrix $AV$ can be computed in $O(nm \log n \log \log n \log m)$ field operations. By Lemma 3.1, to guarantee a high success probability, it is enough to work on a field with $\Theta(n^4)$ elements, so that each field operation can be done in $\tilde{O}(\log n)$ steps [20]. This gives an alternative method to compute $\mathrm{rank}(A)$ of an $m \times n$ matrix in $\tilde{O}(nm \log m (\log n)^2 + m^\omega \log n)$ steps, which is slower than the algorithm in Theorem 1.1 but has similar running time as previous algorithms.

The following is an outline of the dynamic algorithm for computing matrix rank. Given an $m \times n$ matrix $A$ with $m \leq n$, we generate a random $n \times m$ Vandermonde matrix $V$ and we know by Lemma 3.1 that

rank($A$) = rank($AV$) with high probability. We reduce $AV$ to the rank normal form by elementary row and column operations, and maintain the decomposition that $XAVY = D$ where $X$ and $Y$ are $m \times m$ invertible matrices and $D = \begin{pmatrix} I_r & 0 \\ 0 & 0 \end{pmatrix}$. Then rank($A$) = $r$ with high probability. We briefly describe how to maintain the rank under different operations. If we do a rank one update on $A$ (i.e. $A \leftarrow A + uv^T$ where $u$ and $v$ are column vectors), then it corresponds to a rank one update on $D$ and we can bring it back to the rank normal form efficiently. If we add a column to or delete a column from $A$, then we can add a row to or delete a row from $V$ so that rank($AV$) is still equal to rank($A$) with high probability, because of the structure of Vandermonde matrices. If we add a row to or delete a row from $A$, then we can do some rank one updates to maintain the structure of $V$ and rank($AV$) = rank($A$) with high probability. The most interesting case is when $m < n$ is changed to $m > n$ or vice versa. In this case we can change the decomposition of $D = XAVY$ to $D = (XV^{-1})VA(VY)$ and set the new $X$ to be $XV^{-1}$ and the new $Y$ to be $VY$, and this can be implemented efficiently by fast Fourier transform and fast inverse Fourier transform.

**Lemma 3.3.** *Given an $m \times n$ matrix $A$ over a field $F$, there is a data structure that maintains* rank($A$) *supporting the following operations.*

1. *The data structure can be initialized in $O(mn \log m \log n \log \log(m+n) + (\min\{m,n\})^\omega)$ field operations.*

2. rank($A$) *can be updated in $O(mn)$ field operations if a rank one update is performed on $A$.*

3. rank($A$) *can be updated in $O(mn(\log \min\{m,n\})^2)$ field operations if a row or a column is added to or deleted from $A$.*

*The data structure requires space to store $O(mn)$ elements in $F$. The probability of failure in any operation is at most $O(\tilde{n}m^2/|F|)$, where $\tilde{n}$ is the maximum $n$ throughout the updates.*

*Proof.* The data structure stores six matrices $X \in F^{m \times m}$, $A \in F^{m \times n}$, $V \in F^{n \times m}$, $E \in F^{m \times m}$, $Y \in F^{m \times m}$, $D \in F^{m \times m}$. Let $B = AV$ if $m \leq n$ and $B = VA$ if $m > n$. In the following we assume that $m \leq n$, when $m > n$ all the procedures are done in a symmetric manner. We maintain $D = XBY$ with the following properties.

1. $X, Y$ are invertible.

2. $V$ is a Vandermonde matrix, i.e., $V_{ij} = g_i^j$ for some $g_i \in F$.

3. $D$ is a matrix in the form $D = \begin{pmatrix} I_r & 0 \\ 0 & 0 \end{pmatrix}$ where $I_r$ is an $r \times r$ identity matrix.

**Initialization:** We choose $g$ uniformly randomly in $F$ and set $V_{ij} = g^{ij}$. We can reduce $B = AV$ into the rank normal form in $O(m^\omega)$ field operations, and thus obtain $X$ and $Y$ such that $X$ and $Y$ are both invertible and $XBY = \begin{pmatrix} I_r & 0 \\ 0 & 0 \end{pmatrix}$ where $r$ is the rank of $B$ (see e.g. Proposition 16.13 of [7]). This completes the initialization. Note that computing the $i$-th row of $B$ is equivalent to doing a multipoint evaluation of a degree $n$ polynomial with coefficients defined by the $i$-th row of $A$ on the points $g, g^2, \ldots, g^m$. Thus each row can be computed in $O(n \log n \log \log n \log m)$ field operations by Theorem 3.2, and the total cost for computing $B$ is $O(nm \log m \log n \log \log n)$ field operations.

**Rank one update:** Let $A' = A + u'v'^T$ where $u' \in F^{m \times 1}$ and $v' \in F^{n \times 1}$. Then $XA'VY = D + (Xu')(v'^TVY) = D + uv^T$ where $u = Xu'$ and $v^T = v'^TVY$, and $u$ and $v$ can be computed in $O(m^2)$ and $O(nm)$ field operations respectively. And $B' = A'V = AV + u'v'^TV$ can be updated in $O(nm)$ field operations. Suppose $u_i \neq 0$ where $u_i$ is the $i$-th entry of $u$. Let $E_1 = I - (e_i - (1/u_i)u)e_i^T$ where $e_i$ is the $i$-th standard unit vector. Then $E_1$ is invertible and $E_1 XA'VY$ is a sum of a diagonal matrix, at most one nonzero row and at most one nonzero column[1]. Hence we can use $O(m)$ elementary row and column operations to transform $E_1 X A'VY$ into a matrix with each row and column having at most one non-zero entry, where the only nonzero entry is one. This matrix can be further transformed to the rank normal form $D' = \begin{pmatrix} I_{r'} & 0 \\ 0 & 0 \end{pmatrix}$ by using two permutation matrices to permute the rows and columns, where $r'$ is the rank of $XA'VY$. Let $E_2$ be the composition of elementary row operations done, $E_3$ be the composition of elementary column operations done, $P_1$ be the permutation of rows and $P_2$ is the permutation of columns. Then $P_1 E_2 E_1 X A'VY E_3 P_2 = D'$. Note that $X' = P_1 E_2 E_1 X$ and $Y' = Y E_3 P_2$ can be computed in $O(m^2)$ field operations. This is because $E_1$ and $E_2$ are compositions of $O(m)$ elementary operations and each elementary operation acting on an $m \times m$ matrix can be done in $O(m)$ field operations. Also permutations of rows and columns can be done in $O(m^2)$ field operations. Now we have $X'A'VY' = D'$, where $A'$ is updated in $O(nm)$ field operations, and $X'$, $Y'$ and $D'$ are updated in $O(m^2)$ field operations.

**Adding a column or adding a row:** To add a column or a row, we can first add a zero column or a zero row and then do a rank one update. Since we know how to do rank one updates, we restrict our attention to adding a zero column and adding a zero row. Suppose we add a zero column in the end. Then we set $A' = (A, 0)$ and $V'_{n+1,j} = g^{cj}$ where $c$ is the smallest index such that $g^c \neq V_{i,1}$ for $1 \leq i \leq n$. Adding a zero column in the $i$-th column of $A$ is done similarly by adding a new row in the $i$-th row of $V$. Then we maintain that $B = A'V'$ and $D = XBY$. Suppose we add a zero row in the end. Then we set $A' = \begin{pmatrix} A \\ 0 \end{pmatrix}$, and set $V'_{i,m+1} = (V'_{i,1})^{m+1}$ for all $i$. Then we update $B' = A'V' = \begin{pmatrix} B & AV'_{m+1} \\ 0 & 0 \end{pmatrix}$ in $O(nm)$ field operations where $V'_{m+1}$ is the $(m+1)$-th column of $V'$. Note that $\begin{pmatrix} X & 0 \\ 0 & 1 \end{pmatrix} \begin{pmatrix} B & 0 \\ 0 & 0 \end{pmatrix} \begin{pmatrix} Y & 0 \\ 0 & 1 \end{pmatrix} = \begin{pmatrix} D & 0 \\ 0 & 0 \end{pmatrix}$ and the difference between $B'$ and $\begin{pmatrix} B & 0 \\ 0 & 0 \end{pmatrix}$ is a single column, which is a rank one matrix. By the same argument used in rank one update, we can update $X$, $Y$ and $D$ in $O(m^2)$ field operations accordingly. If the zero row is not added at the end, we can first permute the rows so that the added row is at the end, by updating $X$ with $XP$ where $P$ is the corresponding permutation matrix, and then do the above procedure. Then we maintain that $B' = A'V'$ and $D' = X'B'Y'$.

**Deleting a column or deleting a row:** To delete a column or a row, we can do a rank one update to set the column or row to zero, and then delete a zero column or a zero row. So we restrict our attention to deleting a zero column or a zero row. Deleting a zero column is done by deleting the corresponding row in $V$. There is no change to $X, B, Y, D$, and we maintain $B = A'V'$ and $D = XBY$. Suppose we delete a zero row at the end of $A$ to obtain $A'$. Then we delete the last column of $V$ to obtain $V'$. Let $B' = A'V'$. Note that $B'' = \begin{pmatrix} B' & 0 \\ 0 & 0 \end{pmatrix}$ and $B = \begin{pmatrix} B' & AV_m \\ 0 & 0 \end{pmatrix}$ where $V_m$ is the $m$-th column of $V$, and so the difference is a rank one update. So we can find $X''$, $Y''$, and $D''$ such that $X''B''Y'' = D''$ where $D'' = \begin{pmatrix} I_r & 0 \\ 0 & 0 \end{pmatrix}$. Now let $X'$, $Y'$, and $D'$ be obtained by deleting the last row and column of $X''$, $Y''$ and $D''$ respectively. Then $X'B'Y' = D'$, because $D''_{ij} = \sum_{k=1}^{m} \sum_{l=1}^{m} X''_{i,k} B''_{k,l} Y''_{l,j} = \sum_{k=1}^{m-1} \sum_{l=1}^{m-1} X'_{i,k} B'_{k,l} Y'_{l,j} = D'_{i,j}$ for $1 \leq i \leq m-1$ and $1 \leq j \leq m-1$ as $B_{k,l} = 0$ if $k = m$ or $l = m$. Clearly the updates can be done in $O(nm)$ field operations. If the zero row deleted is not at the end, we can first permute the rows so that the deleted row is at the end, by updating $X$ with $XP$ where $P$ is the corresponding permutation matrix, and then do the above procedure.

**Changing representation:** Note that in the above operations we assume $m \leq n$. Some operations require $O(m^2)$ field operations and thus if $m > n$ this is greater than $O(mn)$. Instead we will maintain $B = VA$ and $D = XBY$ when $m > n$. To change the representation, when $m = n$, we rewrite $D = XAVY = (X(V')^{-1})V'A(VY)$, and set $X' = X(V')^{-1}$ and $Y' = VY$ and $V'_{ij} = g^{ij}$ for all $1 \leq i, j \leq n$. Note that $VY$ can be computed in $O(n^2 \log^2 n \log \log n)$ field operations by Theorem 3.2, since it is equivalent to $n$ points evaluation of $n$ degree $n$ polynomials. Moreover, after reseting $V'_{ij} = g^{ij}$, using Theorem 3.4 we can compute $X(V')^{-1}$ in $O(n^2 \log^2 n \log \log n)$ field operations, as it is equivalent to doing $n$ points interpolation $n$ times for each pairs of rows of $X$ and $X(V')^{-1}$. Also $B' = V'A$ can be computed in $O(n^2 \log^2 n \log \log n)$ field operations by the multipoint evaluation algorithm in Theorem 3.2. Therefore, we maintain $B' = V'A$ and $X'B'Y' = D$, and this can be used to support the above operations in a symmetric manner.

---

[1] The details are as follows: $E_1 uv^T = uv^T + (e_i - (1/u_i)u)u_i v^T = u_i e_i v^T$ is a matrix with only one row is non-zero. If $i > r$, then $E_1 X A'VY = E_1(D + uv^T) = E_1 D + u_i e_i v^T = D + u_i e_i v^T$ since $e_i^T D = 0$. If $i \leq r$, then $E_1 X A'VY = E_1 D + u_i e_i v^T = D - (e_i - (1/u_i)u)e_i^T + u_i e_i v^T$ if $i \leq r$ since $e_i^T D = e_i^T$. In either case $E_1 X A'VY$ is a sum of a diagonal matrix, at most one nonzero row and at most one nonzero column.

**Theorem 3.4** ([20] Corollary 10.13). *There is an algorithm that takes $n$ points $(x_i, y_i) \in F^2$ as input and returns a polynomial $f \in F[x]$ with degree less than $n$ which satisfies $f(x_i) = y_i$ for each $i$. The algorithm takes $O(n \log^2 n \log \log n)$ field operations.*

**Error probability:** The rank query will only fail when $\text{rank}(A) \neq \text{rank}(AV)$ at some point, which happens with probability at most $O(\tilde{n} m^2 / |F|)$ by Lemma 3.1. This completes the proof. □

Let $Q$ be an upper bound on the number of updates to the matrix. Then $\tilde{n} \leq n + Q$. By setting $|F| = \Theta((n+Q)^3 m^3)$, then the probability that the algorithm does not make any error in the whole execution is at least $1 - O(1/((n+Q)m))$, while each field operation requires $\tilde{O}(\log((n+Q)m))$ steps. This proves Theorem 1.2.

## 4 Applications

In this section we will show some applications of Theorem 1.1 and Theorem 1.2 to problems in numerical linear algebra, combinatorial optimization, and dynamic data structures. In each subsection we will state the problems, describe the previous work, and present the improvements.

### 4.1 Numerical Linear Algebra

Let $A$ be an $m \times n$ matrix over a field $F$. Let $r = \text{rank}(A)$. The rank-one decomposition of $A$ is to write $A$ as the sum of $r$ rank one matrices. The null space of $A$ is the subspace of vectors for which $Ax = 0$, and the problem is to find a basis of the null space of $A$. The matrix multiplication problem is to compute $AB$ for two $n \times n$ matrices $A$ and $B$. We will show that these problems can be solved faster when $r$ is small. The previous best known algorithms require $\Theta(nmr^{\omega-2})$ field operations, where the bottleneck of these algorithms is in finding a set of $r$ linearly independent columns. Note that previous randomized algorithms for computing $r$ cannot be used to solve these problems, as they do not find a set of $r$ linearly independent columns. In the following we assume that $|F| = \Omega(m+n)$ and $|A| = \Omega(m+n)$.

**Rank-one decomposition:** Without loss of generality we assume $m \leq n$; otherwise we consider $A^T$ instead of $A$. By Theorem 1.1, we can find a set of $r$ independent columns of $A$ in $O((|A| + r^\omega) \log n)$ field operations, with success probability at least $1 - O(\log n / n^{1/3})$. Let $T \subseteq [n]$ be a set of $r$ independent columns, and $S \subseteq [m]$ with $|S| = r$ be the set of rows such that $A_{S,T}$ is of full rank. Again by Theorem 1.1 we can find $S$ in $O(|A| + r^\omega)$ field operations with success probability at least $1 - O(\log n / n^{1/3})$. Now set $B = A_{[m],T}$ and $C = A^{-1}_{S,T} \times A_{S,[n]}$. Then $C_{[r],T} = I_r$ and thus $(BC)_{[m],T} = A_{[m],T}$. Similarly $(BC)_{S,[n]} = A_{S,[n]}$, and thus the entries of $BC$ and $A$ match in the rows of $S$ and also the columns of $T$. Note that both $BC$ and $A$ are of rank $r$, and both $BC_{S,T}$ and $A_{S,T}$ are of full rank. So for any $i \notin S$ and $j \notin T$, $\det(A_{S\cup\{i\}, T\cup\{j\}}) = 0$ and thus $A_{ij}$ is uniquely determined by other entries of $A$. The same applies to $BC$ and thus $A = BC$. Clearly $C$ can be computed in $O(r^{\omega(1,1,\log_r n)})$ field operations. Thus the overall complexity is $O((|A| + r^\omega) \log n + r^{\omega(1,1,\log_r n)})$ field operations.

**Null space:** By the above algorithm for rank-one decomposition, we can find $S \subseteq [m]$, $T \subseteq [n]$, $B \in F^{m \times r}$, and $C \in F^{r \times n}$ such that $A = BC$, $|S| = |T| = r$ and $C_{S,T} = I_r$, with required probability and time complexity. Note that $Ax = 0 \iff BCx = 0 \iff Cx = 0$ since the columns in $B$ are linearly independent. Since $C_{[r],T} = I_r$, we have $Cx = 0 \iff x_T = -C_{[n]-T}x_{[n]-T}$. Thus the entries of $x_{[n]-T}$ can be arbitrarily assigned and then the entries of $x_T$ is uniquely determined. Assume without loss of generality that $T = \{1, \ldots, r\}$. Then a basis $\{b_i\}$ for $i \in [n] - T$ would be $b_i(k) = -C_{k,i}$ for $1 \leq k \leq r$, and then set $b_i(i) = 1$ and set $b_i(j) = 0$ otherwise.

**Matrix multiplication:** Applying the rank-one decomposition algorithm to $A$ to find $A = A_1 A_2$ for some $A_1 \in F^{n \times r}, A_2 \in F^{r \times n}$ in $\tilde{O}(|A| + r^{\omega(1,1,\log_r n)})$ field operations. Now $A_2 B$ can be computed in $O(n^{\omega(\log_n r,1,1)})$ field operations, and so do $A_1(A_2 B)$ since $\omega(1, c, 1) = \omega(c, 1, 1)$ [26]. So the overall complexity is $\tilde{O}(n^{\omega(\log_n r,1,1)})$ field operations.

## 4.2 Graph Matching

Given an undirected graph $G = (V, E)$, the maximum matching problem is to find a set of maximum number of vertex disjoint edges in $G$. The time complexity of the fastest combinatorial algorithms for this problem is $O(\sqrt{\mathsf{opt}} \cdot |E|)$ [32, 42, 21], where $\mathsf{opt}$ denotes the size of a maximum matching.

There is an algebraic formulation for the maximum matching problem proposed by Tutte [41]. Let $V = \{1, \ldots, n\}$ and $x_e$ be a variable for each edge $e$. Let $A$ be an $n \times n$ matrix where $a_{ij} = x_e$ and $a_{ji} = -x_e$ if $e = ij \in E$ and $a_{ij} = a_{ji} = 0$ otherwise. Tutte [41] proved that $G$ has a perfect matching if and only if $A$ is non-singular, and Lovász [30] generalized it to show that $\text{rank}(A) = 2\mathsf{opt}$. Using the Schwartz-Zippel lemma, Lovász [30] also proved that $\text{rank}(A)$ is preserved with high probability, if we substitute non-zero values for the variables $x_e$ from a sufficiently large field, say of size $\Theta(n^2)$. This implies that the size of a maximum matching can be computed in $O(n^\omega)$ field operations, where each field operation can be performed in $O(\log n)$ steps. With additional non-trivial ideas, Mucha and Sankowski [33] and Harvey [22] showed how to also find a maximum matching in $O(n^\omega)$ field operations. This is faster than the combinatorial algorithms when the graph is dense and the $\mathsf{opt}$ is large, for example when $|E| = \Theta(n^2)$ and $\mathsf{opt} = n$ the combinatorial algorithms require $\Theta(n^{2.5})$ steps.

We prove the statement about graph matching in Theorem 1.4. Suppose $k$ is given and the task is to find a matching of size $\min\{k, \mathsf{opt}\}$. Let $k' = 2\min\{k, \mathsf{opt}\}$. We can first use the algorithm in Theorem 1.1 to find a set $S$ of $k'$ linearly independent columns in $A$ in $\tilde{O}(|A| + (k')^\omega) = \tilde{O}(|E| + (k')^\omega)$ field operations, where $|E|$ is the number of edges in $G$. Let $A_{V,S}$ be the $n \times k'$ submatrix formed by these independent columns. We can apply the algorithm in Theorem 1.1 again on $A_{V,S}$ to find a set $R$ of $k'$ linearly independent rows in $A_{V,S}$ in $\tilde{O}(|A_{V,S}| + (k')^\omega) = \tilde{O}(|E| + (k')^\omega)$ field operations. Let $A_{R,S}$ be the $k' \times k'$ submatrix formed by these rows and columns. Consider $A_{R \cup S, R \cup S}$ which is a matrix with size at most $2k' \times 2k'$ and rank at least $k'$. Note that it is the algebraic formulation for the maximum matching problem in $G[R \cup S]$, where $G[R \cup S]$ denotes the induced subgraph on the vertices corresponding to $R \cup S$. And so there is a matching of size $k'/2 = \min\{k, \mathsf{opt}\}$ in $G[R \cup S]$. We can use the algorithm of Mucha and Sankowski [33] or Harvey [22] to find a matching of size $\min\{k, \mathsf{opt}\}$ in $O(k^\omega)$ field operations. Thus the overall complexity is $\tilde{O}(|E| + k^\omega)$ and this proves the statement about graph matching in Theorem 1.4. To find a matching of size $\mathsf{opt}$, we can first use a linear time 2-approximation greedy algorithm to find a matching $M$ of size at least $\mathsf{opt}/2$, and then set $k = 2|M|$ and run the above algorithm.

We mention two problems where this matching result can be applied. One is the maximum subset matching problem considered by Alon and Yuster [4], which asks what is the maximum number of vertices in $S \subseteq V$ that can be matched in a matching of $G$. They proved that this maximum number is equal to $\text{rank}(A_{S,V})$ where $A_{S,V}$ is the submatrix of the Tutte matrix formed by the rows of $S$. Thus we can use Theorem 1.1 to obtain an $\tilde{O}(|\delta(S)| + |S|^\omega)$ algorithm where $|\delta(S)|$ counts the number of edges with one endpoint in $S$ and another endpoint in $V - S$. This improves upon their result which takes $\tilde{O}(|\delta(S)| \cdot |S|^{(\omega-1)/2})$ steps when $|\delta(S)| \geq |S|^{(\omega+1)/2}$. Another is the maximum matching problem in a lopsided bipartite graph $G = (X, Y; E)$ where one side is much larger than the other side [8], that is $|X| \ll |Y|$. In this case $\mathsf{opt} \leq |X|$ and our algorithm can find a maximum matching in $\tilde{O}(|E| + |X|^\omega)$ steps.

## 4.3 Linear Matroid Intersection and Linear Matroid Parity

In the linear matroid intersection problem, we are given two $r \times n$ matrices $M$ and $N$ where the columns in $M$ and $N$ are indexed by $\{1, \ldots, n\}$, and the task is to find a set $S \subseteq \{1, \ldots, n\}$ of maximum size so that the columns in $S$ are linearly independent in both $M$ and $N$. In the linear matroid parity problem, we are given an $r \times 2n$ matrix where the columns are partitioned into $n$ pairs, and the task is to find a maximum cardinality collection of pairs so that the union of the columns of these pairs are linearly independent.

For the linear matroid intersection problem, Gabow and Xu [19] gave a combinatorial algorithm (using fast matrix multiplication) with time complexity $O(nr(\mathsf{opt})^{1/(4-\omega)}) = O(nr(\mathsf{opt})^{0.62})$ when $\omega \approx 2.38$. Harvey [22] gave an algebraic algorithm with time complexity $O(nr^{\omega-1})$, which is faster for any $\mathsf{opt} \geq r^{0.62}$ when $\omega \approx 2.38$. For the linear matroid parity problem, Gabow and Stallmann [18] gave a combinatorial algorithm (using fast matrix multiplication) with time complexity $O(nr^{\omega-1}(\mathsf{opt}))$, and Cheung, Lau and Leung [11] gave an algebraic algorithm with time complexity $\tilde{O}(nr^{\omega-1})$ by extending Harvey's algorithm.

We prove the statement about linear matroid intersection and linear matroid parity in Theorem 1.4. The linear matroid parity problem is a generalization of the linear matroid intersection problem, and any algorithm for the linear matroid parity problem implies an algorithm for the linear matroid intersection with the same time complexity, and so we only consider the linear matroid parity problem in the following. Let $A$ be an $r \times 2n$ matrix where the columns are $\{c_1, c_2, \ldots, c_{2n-1}, c_{2n}\}$ and $(c_{2i-1}, c_{2i})$ is a column pair for $1 \leq i \leq n$. Suppose $k$ is given and the task is to find $\min\{k, \mathsf{opt}\}$ pairs of columns so that the union of the columns of these pairs are linearly independent. We use the algorithm in Theorem 2.5 to compress the matrix $A$ into a $O(k) \times 2n$ matrix $A'$ in $O(|A|)$ field operations, and let the columns of $A'$ be $\{c'_1, c'_2, \ldots, c'_{2n-1}, c'_{2n}\}$. Let $k' = \min\{k, \mathsf{opt}\}$. Assume without loss of generality that the columns in $S = \{c_1, c_2, \ldots, c_{2k'-1}, c_{2k'}\}$ are linearly independent. We claim that the columns in $S' = \{c'_1, c'_2 \ldots, c'_{2k'-1}, c'_{2k'}\}$ are linearly independent with high probability. Consider the submatrix $A_{R,S}$ of $A$ where $R$ is the set of all $r$ rows of $A$. By Theorem 2.5 we have that $\text{rank}(A_{R,S}) = \text{rank}(A'_{R',S'})$ with high probability where $R'$ is the set of all $O(k)$ rows of $A'$. Since $\text{rank}(A_{R,S}) = 2k'$, it implies that $\text{rank}(A'_{R',S'}) = 2k'$ and thus the columns in $S'$ are linearly independent with high probability, proving the claim. Therefore we can apply the algorithm in [11] to solve the matroid parity problem on $A'$, and this can be done in $O(nk^{\omega-1})$ field operations since $A'$ is a $O(k) \times 2n$ matrix. This proves the statement about linear matroid intersection and linear matroid parity in Theorem 1.4.

To find a solution of size $\mathsf{opt}$, we can set $k = 2, 4, 8, \ldots, 2^{\log_2 r}$ and apply the above algorithm until there is no solution of size $k$ or there is a solution of size $r$. A direct implementation of this idea gives an algorithm to find an optimal solution in $O(|A| \log \mathsf{opt} + n(\mathsf{opt})^{\omega-1})$ field operations. We can slightly improve this to $O(\min\{|A| \log \mathsf{opt}, nr\} + n(\mathsf{opt})^{\omega-1})$ field operations by computing the compressed matrices aggregately, but the details are omitted here. Since $\mathsf{opt} \leq r$, our algorithm is faster than the algorithms in [19, 22].

## 4.4 Linear Matroid Union

In the linear matroid union problem, we are given an $r \times n$ matrix $A$ with $r \leq n$, and the task is to find a set of maximum number of disjoint bases, where a basis is a set of maximum number of linearly independent columns, and two bases are disjoint if they do not share any column. For example, the problem of finding a set of maximum number of edge disjoint spanning trees in an undirected graph is a special case of the linear matroid union problem. Let $\mathsf{opt}$ be the maximum number of disjoint bases in $A$, and $b$ be the number of columns in a basis. Cunningham [15] gave a combinatorial algorithm with time complexity $O(nrb(\mathsf{opt}) + nb^2(\mathsf{opt})^2)$.

There is a well known reduction from the linear matroid union problem to the linear matroid intersection problem [38]. Suppose $k$ is given and the task is to find $k$ disjoint bases of $A$ or determine that none exist. Let $M$ be the $kr \times kn$ matrix

$$\begin{pmatrix} A & 0 & \ldots & 0 \\ 0 & A & \ldots & 0 \\ 0 & 0 & \ldots & 0 \\ 0 & 0 & \ldots & A \end{pmatrix},$$

where $0$ denotes the $r \times n$ all zero matrix. Let $N$ be an $n \times kn$ matrix $(I, I, \ldots, I)$ where $I$ is the $n \times n$ identity matrix. Then it can be checked that $A$ has $k$ disjoint bases if and only if the linear matroid intersection problem for $M$ and $N$ has a solution of size $kb$. A direct application of Harvey's algorithm [22] for linear matroid intersection gives an algorithm with time complexity $O((kn) \cdot (kr)^{\omega-1} + (kn) \cdot n^{\omega-1}) = O(nr^{\omega-1}k^\omega + n^\omega k)$.

To do better, we can first reduce the matrix $A$ into a matrix with $kb$ columns before running a linear matroid intersection algorithm, as follows. We can use a compact algebraic formulation for linear matroid intersection [22] where $B = \sum_{i=1}^{kn} x_i \cdot m_i \cdot n_i^T$ where $x_i$ is a random element from a sufficiently large field (say of size $\Theta(n^2)$) and $m_i$ and $n_i$ are the $i$-th column of $M$ and $N$ respectively. For this particular $M$ and $N$, we have

$$B = \begin{pmatrix} A_1 \\ A_2 \\ \vdots \\ A_k \end{pmatrix}$$

where each column of $A_i$ is equal to the corresponding column of $A$ multiplied by an independent random field element. Using the result for linear matroid intersection, it can be shown that if $A$ has $k$ disjoint bases, then $B$ has rank $kb$ with high probability. Furthermore, if we find $kb$ linearly independent columns in $B$, then the corresponding columns in $A$ can be partitioned into $k$ disjoint bases. So, one can first find a set of $kb$ linearly independent columns in $B$ in $O(n(kr)^{\omega-1})$ field operations by Gaussian elimination (or conclude that there are no $k$ disjoint bases if none exist), and then delete the other columns and consider the linear matroid union problem for the $r \times kb$ submatrix of $A$. Then we can run the linear matroid intersection algorithm [22] to find the $k$ disjoint bases in $O((kb)r^{\omega-1}k^\omega + (kb)^\omega k) = O(r^{\omega-1}bk^{\omega+1})$ field operations by using $n = kb$ and $b \leq r$. This results in an $O(nr^{\omega-1}k^{\omega-1} + r^{\omega-1}bk^{\omega+1})$ algebraic algorithm for the linear matroid union problem using existing techniques, although it was not explicitly stated in the literature.

We are now ready to prove the statement about linear matroid union in Theorem 1.4. First we use the compression algorithm in Theorem 2.5 to reduce $A$ to a $O(b) \times n$ matrix $A'$ with $|A'| = O(|A|)$ in $O(|A|)$ field operations. By the same argument used in linear matroid parity, we can show that if $A$ has $k$ disjoint bases, then $A'$ has the same $k$ disjoint bases with high probability. We construct the $O(kb) \times n$ matrix $B'$ as in the previous paragraph in $O(k|A'|) = O(k|A|)$ field operations since $|A'| = O(|A|)$. Then we use the algorithm in Theorem 1.1 to find $kb$ linearly independent columns in $B'$ in $\tilde{O}(k|A| + (kb)^\omega)$ field operations since $kb \leq n$ (or conclude that there are no $k$ disjoint bases if none exist). As stated in the previous paragraph, the corresponding $kb$ columns in $A'$ can be partitioned into $k$ disjoint bases with high probability. So we delete other columns and only consider the $O(b) \times kb$ submatrix $A''$ of $A'$. Now we have reduced the linear matroid parity problem for an $r \times n$ matrix $A$ to the linear matroid union problem for a $O(b) \times kb$ matrix $A''$. We can run Harvey's linear matroid intersection algorithm using the above reduction to find the $k$ disjoint bases in $O((kb)b^{\omega-1}k^\omega + (kb)^\omega k) = O(b^\omega k^{\omega+1})$ field operations by putting $n = kb$ and $r = O(b)$. Alternatively, we can use Cunningham's algorithm to find the $k$ disjoint bases in $O((kb)b^2k + (kb)b^2k^2) = O(b^3k^3)$ field operations by putting $n = kb$ and $r = O(b)$. Therefore the total complexity is $\tilde{O}(k|A| + \min\{b^\omega k^{\omega+1}, b^3k^3\})$ field operations where $|A| \leq nr$, proving the statement about linear matroid union in Theorem 1.4. To find the maximum number of disjoint bases, we can use doubling ($k = 2, 4, 8, \ldots, 2^{\log \text{opt}}$) and then binary search, and apply the above algorithm as in linear matroid parity, and obtain an algorithm with time complexity $\tilde{O}(\log \mathsf{opt}(nr(\mathsf{opt}) + \min\{b^\omega(\mathsf{opt})^{\omega+1}, b^3(\mathsf{opt})^3\})$ field operations. Ignoring polylog factors, this is faster than the previous algorithms for any values of $r, b, \mathsf{opt}, n$.

## 4.5 Dynamic Edge Connectivities

In this section we show that the dynamic matrix rank algorithm in Theorem 1.2 can be applied to obtain an efficient dynamic algorithm for computing all pairs edge connectivities in a simple directed graph $G = (V, E)$, supporting the operations of adding and deleting edges. The $s$-$t$ edge connectivity is defined as the size of a minimum $s$-$t$ cut, or equivalently the number of edge disjoint directed paths from $s$ to $t$.

We will use a recent algebraic formulation that relates edge connectivities to matrix ranks [12]. Construct an $|E| \times |E|$ matrix $M$ as follows:

$$M_{i,j} = \begin{cases} x_{i,j} & \text{if the head of } e_i \text{ is equal to the tail of } e_j \\ -1 & \text{if } i = j \\ 0 & \text{otherwise} \end{cases}$$

The matrix has the following properties:

**Theorem 4.1** ([12]). *The $s$-$t$ edge connectivity is equal to the rank of the submatrix $M^{-1}{}_{\delta^{out}(s), \delta^{in}(t)}$, where $\delta^{in}(v)$ and $\delta^{out}(v)$ are the set of incoming and outgoing edges of $v$ respectively. In addition, if we substitute random values to $x_{i,j}$ from a field $F$, the claim still holds with probability at least $1 - O(|E|^2/|F|)$.*

This formulation implies an $O(|E|^\omega)$ time algorithm for computing all pairs edge connectivities for simple directed graphs. We are going to show that using the dynamic matrix rank algorithm, we can support each adding and deleting edge operation in $\tilde{O}(|E|^2)$ time, by maintaining the ranks of all the submatrices $(M^{-1})_{\delta^{out}(s), \delta^{in}(t)}$ dynamically.

First we consider the case of adding an edge. Let $G$ be the original graph, and $\tilde{G}$ be the graph with an edge added to $G$. Let $M$ and $\tilde{M}$ be the edge connectivity matrix formulation for $G$ and $\tilde{G}$ respectively. Observe that $\tilde{M}$ is obtained from $M$ by adding one extra row and one extra column at the end. We will maintain $M^{-1}$ and the ranks of its submatrices, by first adding a trivial row and column to $M$, and then fill in the required entries. Let

$$M' = \begin{pmatrix} M & 0 \\ 0 & -1 \end{pmatrix} \text{ and } (M')^{-1} = \begin{pmatrix} M^{-1} & 0 \\ 0 & -1 \end{pmatrix}.$$

Since we only have to modify the last row and last column of $M'$ to get $\tilde{M}$, we can write $\tilde{M} = M' + UV^T$ for two $(|E|+1) \times 2$ matrices $U = (\vec{e}, c)$ and $V = (r, \vec{e})$, where $c$ is the new column with $|E| + 1$ entries, $r^T$ is the new row with $|E| + 1$ entries, and $\vec{e}$ be the column vector with the first $|E|$ entries to be zero and the last entry to be one. The following result shows that such a low rank update can be computed efficiently.

**Theorem 4.2** (Sherman-Morrison-Woodbury [44]). *Suppose that matrices $M$ and $M + UV^T$ are both non-singular, then $(M + UV^T)^{-1} = M^{-1} - M^{-1}U(I + V^T M^{-1} U)^{-1} V^T M^{-1}$.*

By the theorem $\tilde{M}^{-1}$ is also a rank-2 update to $(M')^{-1}$, and $\tilde{M}^{-1}$ can be obtained from $M^{-1}$ in $O(|E|^2)$ time since $U$ and $V$ are $(|E|+1) \times 2$ matrices. Similarly, any submatrix $(\tilde{M}^{-1})_{\delta^{out}(s)\delta^{in}(t)}$ can be obtained by a rank-2 update to $((M')^{-1})_{\delta^{out}(s), \delta^{in}(t)}$, and $((M')^{-1})_{\delta^{out}(s), \delta^{in}(t)}$ can be obtained by adding at most one row and one column to $(M^{-1})_{\delta^{out}(s), \delta^{in}(t)}$. Since both operations are supported by our dynamic matrix rank algorithm in $\tilde{O}(|\delta^{out}(s)||\delta^{in}(t)|)$ field operations, we can maintain $\text{rank}((\tilde{M}^{-1})_{\delta^{out}(s)\delta^{in}(t)})$ and thus the $s$-$t$ edge connectivity between any pair of vertices $s$ and $t$ in $\tilde{O}(|\delta^{out}(s)||\delta^{in}(t)|)$ field operations after an edge is added. Thus we can maintain all pairs edge connectivities in $\tilde{O}(\sum_{s,t \in V} |\delta^{out}(s)||\delta^{in}(t)|) = \tilde{O}(|E|^2)$ field operations. Let $Q$ be an upper bound on the number of edge updates throughout the whole algorithm. For one pair, by the result in Section 3, the probability that the algorithm makes some mistake during the whole algorithm is at most $O(1/(Q|E|^3))$, if the field size is $\Theta(|E|^5 Q^3)$. Therefore, the probability that the algorithm makes a mistake for some pair during the whole algorithm is at most $O(1/(Q|E|))$. Therefore, each field operation can be done in $\tilde{O}(\log(|E|Q))$ steps.

The case of deleting an edge is almost the same. Assume we are deleting the edge that correspond to the last row and column of $M$. We first write zero to all the entries of that row and column except keeping $M_{|E|,|E|} = -1$, and then we delete the last row and column. These two steps correspond to a rank-2 update followed by a row and column deletion on $M^{-1}$ using the dynamic matrix rank algorithm. This is just the reverse process for adding an edge. By the same argument as above, the new inverse and the ranks of all the required submatrices can be updated in $O(|E|^2)$ field operations, where each field operation can be done in $\tilde{O}(\log(|E|Q))$ steps. Note that we just require $O(|E|^2 \log(|E|Q))$ space to store the inverse of $M$. This proves Theorem 1.5.

# Acknowledgement

The research is supported by Hong Kong RGC grant 413411. We thank Arne Storjohann for pointing out the previous results on computing matrix rank in [29, 9], Ankur Moitra for asking the question on dynamic graph connectivity, David Woodruff for pointing out the references on random projection and fast Johnson Lindenstrauss transform, and Nick Harvey and Mohit Singh for useful comments on an earlier draft of this paper.

# References

[1] R. Ahlswede, N. Cai, S.R. Li, and R.W. Yeung. *Network Information Flow*. IEEE Transactions on Information Theory, 46(4):1204–1216, 2000.

[2] N. Ailon, B. Chazelle. *Approximate Nearest Neighbors and the Fast Johnson-Lindenstrauss Transform*. Proceedings of the 38th Annual ACM Symposium on Theory of Computing (STOC), 557–563, 2006.

[3] N. Ailon, E. Liberty. *An Almost Optimal Unrestricted Fast Johnson-Lindenstrauss Transform*. Proceddings of the 22nd ACM-SIAM Symposium on Discrete Algorithms (SODA), 185–191, 2011.

[4] N. Alon, R. Yuster. *Fast Algorithms for Maximum Subset Matching and All-Pairs Shortest Paths in Graphs with a (Not So) Small Vertex Cover*. Proceedings of the 15th Annual European Symposium on Algorithms (ESA), 175–186, 2007.

[5] A. Bhalgat, R. Hariharan, T. Kavitha, and D. Panigrahi. *An $\tilde{O}(mn)$ Gomory-Hu Tree Construction Algorithm for Unweighted Graphs*. Proceedings of the 39th Annual ACM Symposium on Theory of Computing (STOC), 605–614, 2007.

[6] J.R. Bunch, J.E. Hopcroft. *Triangular Factorization and Inversion by Fast Matrix Multiplication*. Mathematics of Computation, 28(125):231–236, 1974.

[7] P. Bürgisser, M. Clausen, M.A. Shokrollahi. Algebraic Complexity Theory, Springer, 1997.

[8] D. Charles, M. Chickering, N.R. Devanur, K. Jain, M. Sanghi. *Fast Algorithms for Finding Matchings in Lopsided Bipartite Graphs with Applications to Display Ads*. Proceedings of the ACM Conference on Electronic Commerce, 121–128, 2010.

[9] L. Chen, W. Eberly, E. Kaltofen, B.D. Saunders, W.J. Turner, G. Villard. *Efficient Matrix Preconditioners for Black Box Linear Algebra.* Linear Algebra and its Applications, 343-344:119–146, 2002.

[10] J. Cheriyan. *Randomized $\tilde{O}(M(|V|))$ Algorithms for Problems in Matching Theory.* SIAM Journal on Computing, 26(6):1635–1655, 1997.

[11] H.Y. Cheung, L.C. Lau, K.M. Leung. *Algebraic Algorithms for Linear Matroid Parity Problems.* Proceedings of the 22nd ACM-SIAM Symposium on Discrete Algorithms (SODA), 1366–1382, 2011.

[12] H.Y. Cheung, L.C. Lau, K.M. Leung. *Graph Connectivities, Network Coding, and Expander Graphs.* Proceedings of the 52nd Annual IEEE Symposium on Foundations of Computer Science (FOCS), 2011.

[13] D. Coppersmith. *Rectangular Matrix Multiplication Revisited.* Journal of Complexity 13, 42–49, 1997.

[14] D. Coppersmith, S. Winograd. *Matrix Multiplication via Arithmetic Progressions.* Journal of Symbolic Computation, 9:251-280, 1990.

[15] W.H. Cunningham. *Improved Bounds for Matroid Partition and Intersection Algorithms.* SIAM Journal on Computing, 15(4):948–957, 1986.

[16] G.S. Frandsen, P.F. Frandsen. *Dynamic Matrix Rank.* Proceedings of the 33rd International Colloquium on Automata, Language and Programming (ICALP), 395–406, 2006.

[17] O. Gabber, Z. Galil. *Explicit Constructions of Linear-Sized Superconcentrators.* Journal of Computer and System Sciences, 22:407–420, 1981.

[18] H.N. Gabow, M. Stallmann. *An Augmenting Path Algorithm for Linear Matroid Parity.* Combinatorica, 6:123-150, 1986.

[19] H.N. Gabow, Y. Xu. *Efficient Theoretic and Practical Algorithms for Linear Matroid Intersection Problems.* Journal of Computer and System Sciences, 53:129–147, 1996.

[20] J. von zur Gathen, J. Gerhard. Modern Computer Algebra. Cambridge University Press, 2nd edition, 2003.

[21] A.V. Goldberg, A.V. Karzanov. *Maximum Skew-symmetric Flows and Matchings.* Mathematical Programming, 100(3):537–568, 2004.

[22] N. Harvey. *Algebraic Algorithms for Matching and Matroid Problems.* SIAM Journal on Computing, 39:679-702, 2009.

[23] T. Ho, M. Médard, R. Koetter, D.R. Karger, M. Effros, J. Shi, B. Leong. *A Random Linear Network Coding Approach to Multicast.* IEEE Transactions on Information Theory 52, 4413–4430, 2006.

[24] J. Holm, K. de Lichtenberg, M. Thorup. *Poly-logarithmic Deterministic Fully-Dynamic Algorithms for Connectivity, Minimum spanning tree, 2-edge, and Biconnectivity.* Journal of the ACM, 48(4), 2001.

[25] S. Hoory, N. Linial, A. Wigderson. *Expander Graphs and their Applications.* Bulletin (New series) of the American Mathematical Society, 43:439-561, 2006.

[26] X. Huang, V. Pan. *Fast Rectangular Matrix Multiplication and Applications.* Journal of Complexity, 14(2): 257-299, 1998.

[27] O.H. Ibarra, S. Moran, R. Hui. *A Generalization of the Fast LUP Matrix Decomposition Algorithm and Applications.* Journal of Algorithms, 3(1):45–56, 1982.

[28] A.W. Ingleton, M.J. Piff. *Gammoids and Transversal Matroids*. J. Combin. Theory Ser. B, 15, 51–68, 1973.

[29] E. Kaltofen, B.D. Saunders. *On Wiedemann's Method of Solving Sparse Linear Systems*. In Proceedings of the 9th International Symposium, on Applied Algebra, Algebraic Algorithms and Error-Correcting Codes (AAECC-9), 29–38, 1991.

[30] L. Lovász. *On Determinants, Matchings and Random Algorithms*. In L. Budach, editor, Fundamentals of Computation Theory, FCT 79, 565–574. Akademie-Verlag, Berlin, 1979.

[31] L. Lovász. *Matroid Matching and Some Applications*. Journal of Combinatorial Theory Series B, 28:208–236, 1980.

[32] S. Micali, V.V. Vazirani. *An $O(\sqrt{|V|}|E|)$ Algorithm for Finding Maximum Matching in General Graphs*. In Proceedings of the 21st Annual IEEE Symposium on Foundations of Computer Science (FOCS), 17–27, 1980.

[33] M. Mucha, P. Sankowski. *Maximum matchings via Gaussian elimination*. Proceedings of the 45th Annual IEEE Symposium on Foundations of Computer Science (FOCS), 248–255, 2004.

[34] P. Sankowski. *Dynamic Transitive Closure via Dynamic Matrix Inverse*. Proceedings of the 45th Annual IEEE Symposium on Foundations of Computer Science (FOCS), 509–517, 2004.

[35] P. Sankowski. *Faster Dynamic Matchings and Vertex Connectivity*. Proceedings of the 18th Annual ACM-SIAM Symposium on Discrete Algorithms (SODA), 118–126, 2007.

[36] T. Sarlós. *Improved Approximation Algorithms for Large Matrices via Random Projections*. Proceedings of the 47th Annual IEEE Symposium on Foundations of Computer Science (FOCS), 143-152, 2006.

[37] D. Saunders, A. Storjohann, G. Villard. *Matrix Rank Certification*. Electronic Journal of Linear Algebra, 11:16–23, 2004.

[38] A. Schrijver. Combinatorial Optimization. Springer, 2003.

[39] A. Storjohann. *Integer Matrix Rank Certification*. Proceedings of the International Symposium on Symbolic and Algebraic Computation (ISSAC), 2009.

[40] L.N. Trefethen, D. Bau III. Numerical Linear Algebra, SIAM, 1997.

[41] W.T. Tutte. *The factorization of linear graphs*. Journal of the London Mathematical Society, 22:107–111, 1947.

[42] V.V. Vazirani. *A Theory of Alternating Paths and Blossoms for Proving Correctness of the $O(\sqrt{|V|}|E|)$ General Graph Matching Algorithm*. In Proceedings of the 1st Integer Programming and Combinatorial Optimization Conference (IPCO), 509–530, 1990.

[43] D.H. Wiedemann. *Solving Sparse Linear Equations Over Finite Fields*. IEEE Transactions on Information Theory 32(1), 54–62, 1986.

[44] M.A. Woodbury. *Inverting Modified Matrices*. Memorandum Report 42, Statistical Research Group, Princeton University, 1950.

# A  Matrix Rank Algorithm by Superconcentrators

In this section we present an algorithm to compute the rank of an $m \times n$ matrix with $m \leq n$ in $O(mn + m^\omega)$ field operations using a superconcentrator.

**Definition A.1** (Superconcentrator). *A superconcentrator is a directed graph $G = (V, E)$ with two given sets $I \subseteq V$ and $O \subseteq V$ with $|I| = |O| = n$, such that for any subsets $S \subseteq I$ and $T \subseteq O$ with $|S| = |T| = k$, there are $|S| = |T|$ vertex disjoint paths from $S$ to $T$.*

There exist superconcentrators with the following properties: (1) there are $O(n)$ vertices and $O(n)$ edges, (2) the indegrees and the outdegrees are bounded by a constant, and (3) the graph is acyclic. Moreover, the construction of such a superconcentrator can be done in linear time [25]. A superconcentrator can be used to obtain an efficient compression algorithm.

**Lemma A.2.** *Given an $m \times n$ matrix $A$ over a field $F$ and an integer $k \leq \min\{n, m\}$, there is an algorithm to construct an $m \times k$ matrix $B$ over $F$ in $O(nm)$ field operations, such that $\operatorname{rank}(B) = \min\{\operatorname{rank}(A), k\}$ with probability at least $1 - nm/|F|$.*

*Proof.* To construct the matrix $B$, we first construct a superconcentrator $G = (V, E)$ with $|I| = |O| = n$ in linear time. Add a source vertex $s$ and add edges from $s$ to each node in $I$ in $G$. We call these edges input edges. Add a sink vertex $t$ and edges from each node in $O$ to $t$ in $G$. We call these edges output edges. Now we associate each edge $e \in E$ with an $m$-dimensional vector $\vec{v}_e$ in $F$. The $n$ vectors $\vec{v}_{su}$ corresponding to the input edges are set to be the column vectors of $A$. Next, for each node $u \in V - \{s, t\}$, for each incoming edge $xu$ and each outgoing edge $uy$, associate the pair of edges $(xu, uy)$ with a random coefficient $c_{(xu,uy)} \in F$. Now we process the nodes of $G$ in a topological order. For each node $u \in V - \{s, t\}$, set each vector associated with the outgoing edge $\vec{v}_{uy}$ to be $\sum_{xu \in E} c_{(xu,uy)} \vec{v}_{xu}$. Finally we choose the vectors associated with the first $k$ output edges to be the column vectors of $B$ and output the matrix $B$.

Since the indegrees and the outdegrees are bounded by a constant, the number of field operations required to process one node in $G$ is $O(m)$. Therefore the algorithm takes $O(nm)$ field operations.

We analyze the probability that $\operatorname{rank}(B) = \min\{\operatorname{rank}(A), k\}$. Let $k' = \min\{\operatorname{rank}(A), k\}$. Clearly $\operatorname{rank}(B) \leq k'$ since the column space of $B$ is spanned by the column space of $A$ and $B$ has only $k$ columns. So we only need to show the other direction. Assume without loss of generality that $A_{[k'],[k']}$ is of full rank. By the property of the superconcentrator $G$, there exists $k'$ vertex disjoint paths from the first $k'$ input nodes to the first $k'$ output nodes. Set $c_{(xu,uy)} = 1$ if the edges $xu$ and $uy$ belongs to one of the paths, and $c_{(xu,uy)} = 0$ otherwise. Then all edges in the path containing the $j$-th input node is associated with the $j$-th column vector of $A$. Thus the $k' \times k'$ submatrix $B_{[k'],[k']}$ of the output matrix $B$ is the same as $A$, up to permutation of columns, and thus it is of rank $k'$. Therefore we can conclude that with non-zero probability the above algorithm outputs a matrix $B$ with $\operatorname{rank}(B) \geq \operatorname{rank}(B_{[k'],[k']}) = k'$. Finally, note that for a fixed input $A$, each entry in the output matrix $B$ is a multivariate polynomial with total degree $O(n)$ (which is the length of a longest path in $G$) with variables $c_{(e1,e2)}$. Therefore the determinant of the first $k'$ columns of $B$ is a multivariate polynomial of total degree $O(nk') = O(nm)$. By the Schwartz-Zippel lemma, if we substitute the variables with random elements in $F$, the probability that the determinant of $B_{[k'],[k']}$ is nonzero and thus $\operatorname{rank}(B) \geq k'$ is at least $1 - O(nm/|F|)$. □

Compared with the algorithm in Theorem 1.1 using magical graphs, this algorithm has the advantage the compressed matrix is of size $k \times k$ rather than of size $O(k) \times O(k)$.

Also we can obtain an algorithm to compute $r = \operatorname{rank}(A)$ in $O(mn + r^\omega)$ field operations as follows. First we apply Lemma A.2 on $A$ and get an $m \times n$ output matrix $B$, and then apply Lemma A.2 on $B^T$ and an get an $n \times m$ output matrix $C$. By Lemma A.2, the resulting matrix $C$ has the property that the rank of any $k \times k$ submatrix is equal to $\min\{\text{rank}(A), k\}$ with probability at least $1 - O(nm/|F|)$. Therefore, to compute $\text{rank}(A)$, one can set $k = 2, 4, 8, \ldots$ and compute the rank of any $k \times k$ matrix of $C$ until the returned rank is less than $k$. The total complexity of this algorithm is only $O(mn + r^\omega)$ field operations where $r = \text{rank}(A)$, which is slightly faster than the $O(|A| \log r + r^\omega)$ algorithm stated in Theorem 2.6.